\documentclass[conference]{IEEEtran}
\IEEEoverridecommandlockouts
\usepackage{cite}
\usepackage{amsmath,amssymb,amsfonts}
\usepackage{algorithmic}
\usepackage{graphicx}
\usepackage{textcomp}
\usepackage{xcolor}
\def\BibTeX{{\rm B\kern-.05em{\sc i\kern-.025em b}\kern-.08em
    T\kern-.1667em\lower.7ex\hbox{E}\kern-.125emX}}

\usepackage{comment}
\usepackage{url}
\usepackage{listings}
\usepackage{booktabs}
\usepackage{subfig}

\begin{document}

\title{Offline vs Real-Time Validation and Performance Assessment of Heimdall Detection Pipeline for Fast Radio Bursts\\
}

\author{
\IEEEauthorblockN{
Valentina Cesare\IEEEauthorrefmark{1},
Giovanni Naldi\IEEEauthorrefmark{1},
Francesco Fiori\IEEEauthorrefmark{1},
Andrea Geminardi\IEEEauthorrefmark{2}, \IEEEauthorrefmark{3}, \IEEEauthorrefmark{4}\\
Adrian De Barro\IEEEauthorrefmark{5},
Alessio Magro\IEEEauthorrefmark{5},
and
Hayley Camilleri\IEEEauthorrefmark{5}}
\\
\IEEEauthorblockA{
\IEEEauthorrefmark{1}IRA, Medicina Radio Astronomical Station, INAF, Medicina (BO), Italy, Email: {name.lastname}@inaf.it\\
\IEEEauthorrefmark{2}IUSS - School for Advanced Studies, Pavia (PV), Italy, Email: andrea.geminardi@iusspavia.it\\
\IEEEauthorrefmark{3}Department of Physics, University of Trento, Povo (TN), Italy\\
\IEEEauthorrefmark{4}Astronomical Observatory of Cagliari, INAF, Selargius (CA), Italy, Email: andrea.geminardi@inaf.it\\
\IEEEauthorrefmark{5}Institute of Space Sciences and Astronomy, University of Malta, Msida, Malta, \\ Email: \{adrian.debarro, alessio.magro, hayley.camilleri.19\}@um.edu.mt}}

\maketitle

\begin{abstract}
Data rates from current radio telescopes are of $\sim$ $O(10)$~PB/year and we expect a growth of about an order of magnitude for future radio telescopes, such as the Square Kilometer Array (SKA). Coping with these data flows requires a paradigm change in data management: building pipelines with intense I/O between their modules (offline mode) is becoming unsustainable, leading to a transition to a real-time paradigm, where data are moved to RAM, e.g., through PSRDADA ring buffers. Our objective is to obtain a compact pipeline to efficiently manage data in the real-time flow, while minimizing high performance computing resources. We employ Heimdall, an open source detection pipeline that performs incoherent de-dispersion, to detect single pulses in radio astronomy data, such as the so called fast radio bursts (FRBs), astrophysical signals whose origin has not been completely explained. Heimdall currently processes a single beam on a single graphics processing unit  (GPU) and can operate both in offline and real-time modes, reading in input either a SIGPROC filterbank file or a PSRDADA ring buffer. We run a test pipeline with data collected from 16 cylinders of the North-South arm of the Northern Cross radio telescope near Medicina (Bologna, Italy), both in offline and real-time modes, to compare the obtained results and performance. To corroborate these tests, we employ different input data and explore the problem's parameter space. PSRDADA and SIGPROC results generally agree within $3\%$ for the most important quantities characterizing a FRB (time of arrival and dispersion measure) and within $10\%$ for the signal-to-noise ratios. Execution times in PSRDADA mode stay in the real-time flow for most runs from data containing true FRBs but they are still generally comparable with SIGPROC mode in this testbed setup. This test configuration provides an important preliminary comparison in terms of accuracy and performance: this work is functional and preparatory for the next integration of the beamformer with the Heimdall pipeline, having the same logical structure as the test pipeline, work already in progress. 
\end{abstract}


\section{Introduction}
\label{sec:Intro}

In recent years, technological evolution has determined an exponential growth of data rates, 
entering in the so called ``Big Data era''. 
Modern radio telescopes, such as MeerKAT~\cite{Jonas_MeerKAT_2016}, ASKAP~\cite{Johnston_ASKAP_2007}, MWA~\cite{Tingay_MWA_2013}, LOFAR~\cite{Rottgering_LOFAR_2003}, and CHIME~\cite{Amiri_CHIME_2018} generate raw data flows of $\sim$100~Gb/s-100~Tb/s\footnote{\url{https://www.mwatelescope.org/telescope/data/}; \url{https://chime-experiment.ca/en/instrument/}; \url{https://www.csiro.au/en/about/facilities-collections/ATNF/ASKAP-radio-telescope}} and scientific data flows toward data centers of $\sim$PB/year or $O(10)$~PB/year (e.g., ASKAP, 75 PB/year~\cite{Chapman_ASKAP_2014}). These rates are expected to increase with next-generation radio telescopes, having a growing number of antennas, such as CHORD~\cite{Vanderlinde_CHORD_2019-2020} and the 
Square Kilometer Array (SKA)~\cite{Hall_SKA_2008}\footnote{\url{https://www.skao.int/en}}. In SKA, the rates of raw data from SKA-Low (Australia) and SKA-Mid (South Africa) telescopes are estimated to be of 2~Pb/s and 20~Tb/s, respectively~\cite{Fenech_SKA_Big_Data_2024}, the rates from Central Signal Processor (CSP) to Science Data Processor (SDP) to be of $7.4$~PB/s and $8.9$~Tb/s~\cite{Garrido_SKA_Big_Data_2026}, and the total scientific data rate 
from the two instruments to the SKA Regional Centers network to be of $\sim$700~PB/year~\cite{Fenech_SKA_Big_Data_2024,Garrido_SKA_Big_Data_2026}.

These data rates impose a paradigm change on the hardware, firmware, software, and data management sides. Currently, radio astronomical analysis is performed with pipelines ported to accelerators~\cite{Van_Tonder_PhD_2014}, where raw data processing, facing higher throughput, 
is executed on low-level hardware, such as field programmable gate array (FPGA)-based boards (e.g.,~\cite{Naldi_FPGA_2017,van_der_Byl_2021}), 
and scientific data processing, facing lower throughput and higher numerical complexity, 
is executed on high-level hardware, such as graphics processing unit (GPU), on a high performance computing (HPC) cluster  (e.g.,~\cite{Barsdell_2012,Magro_2015,Malenta_2020}), due to the different architectural features of the two accelerators~\cite{Van_Tonder_PhD_2014,Hosseinabady_FPGA_GPU_2019,Hennessy_and_Patterson_2025,Wen-Mei_2026}. Radio astronomical pipelines follow a modular structure, where intense I/O on disk between different modules represents one of the greatest issues. Data saved to disk are successively processed by computational infrastructures, in an ``offline'' mode. 
The steep rise of data rates, 
makes this approach unfeasible, both for performance degradation, due to I/O bottleneck, and for storage limitations. This is further critical in time domain astrophysics for the search of radio transients, 
such as Fast Radio Bursts (FRBs)~\cite{Lorimer_2007}, bright extragalactic radio signals with unpredictable behavior, 
arriving from random directions of the sky, with sources difficult to locate  (e.g.,~\cite{Scott_FRB_2025}). The inability to save all raw and uncompressed data to disk would inevitably lead to a loss of essential scientific information, needing 
a transition 
from an offline to a real-time approach. Real-time data processing, with a continuous data flow in RAM, saving to disk only scientifically relevant information, allows to keep up with data acquisition rates, while accessing to raw and uncompressed data~\cite{Barsdell_PhD_Thesis_2012}.

A relevant example of scientific data processing pipeline is given by Heimdall\footnote{\url{https://sourceforge.net/projects/heimdall-astro/}}, an open source and GPU-accelerated pipeline for radio transients detection. Heimdall is built to operate both in offline and real-time modes, reading in input either a SIGPROC filterbank file~\cite{Lorimer_SIGPROC_2011}\footnote{\url{https://sigproc.sourceforge.net}} 
or a PSRDADA ring buffer, respectively. The latter is a circular data structure in RAM to handle continuous data flows in memory with very low latency 
implemented with PSRDADA~\cite{van_Straten_PSRDADA_2021}\footnote{\url{https://psrdada.sourceforge.net}}, an open source HPC software project to support the development of data acquisition and distributed analysis systems, originally developed for radio astronomy targeting pulsar processing.
PSRDADA provides a dual advantage: (1) it provides a ring buffer in shared memory, to manage very high throughput in an optimal way; (2) it is widely employed in several pipelines developed for different radio telescopes (e.g., Zhang, et al.~\cite{Zhang_PSRDADA_2026} employ it for URANUS pipeline to be deployed for the ultrawideband low-frequency receiving system of the QiTai radio telescope), fostering 
portability within the radio astronomical community. Portability is an increasing necessity to cope with Big Data production also in other contexts~\cite{Pennycook_Performance_Portability_2019,Malenza_Performance_Portability_2024_a,Malenza_Performance_Portability_2024_b,Malenza_Performance_Portability_2025}. Another well-known and noteworthy framework for real-time and high-throughput data processing in radio astronomy is Bifrost~\cite{Cranmer_Bifrost_2017}.

We employ Heimdall that processes data acquired by the Northern Cross radio telescope, located near Medicina (Bologna, Italy) and operated by the Istituto Nazionale di Astrofisica (INAF). This instrument is a T-shaped radio interferometer with two arms along North-South (NS) and East-West (EW) directions, dedicated to FRB and space debris detections, operating at a central frequency of $408$~MHz over a bandwidth of $14.81$~MHz.  The telescope features a total collecting area of $\sim$$30000$~m$^2$ and a wide Field of View (FoV), which makes it particularly suitable for FRB searches~\cite{Locatelli_Northern_Cross_2020,Trudu_Northern_Cross_2022,Pelliciari_Northern_Cross_2023,Pelliciari_Northern_Cross_2024,Geminardi_Northern_Cross_2025,De_Barro_2025}. For this work, we take data only from 16 cylinders over 64 (1/4) of NS arm, corresponding to an effective area of $\sim$$2000$~m$^2$ (Section~\ref{sec:Tests_Files}). Heimdall is widely employed in the radio astronomical community, making the relevance of this work not solely confined to this analysis. For example, Pei, et al.~\cite{Pei_Rel_Works_2025} use Heimdall for pulsar and radio transients searches with the QiTai radio telescope and Houben, et al.~\cite{Houben_Rel_Works_2026} for single-pulse searches within the northern part of the HTRU survey collected with the $100$~m Effelsberg Radio Telescope.

In this work, we present a compact pipeline to efficiently manage data in the real-time flow with minimum employment of HPC resources. The purpose
is twofold: 
(1) we aim to accurately validate PSRDADA mode against the more consolidated SIGPROC mode, comparing results obtained from the same input dataset; 
(2) we aim to compare the performances obtained with the two modes in relation with data acquisition times and verify possible performance boosts of real-time mode over offline mode. 
These tests are corroborated by exploring over 
different input data and parameters, 
for a robust determination of scientific results in an efficient way, compelling need for future radio telescopes. The adopted pipeline is defined in a test configuration which follows the same logical structure as the future and complete pipeline that will be deployed in real-time to conduct Northern Cross radio telescope observations, making these tests an essential basis.

Section~\ref{sec:Pipeline} presents the complete pipeline with its offline and real-time modes; Section~\ref{sec:Tests} focuses on the testbed setup, providing details on the software, employed data, and hardware sides; Section~\ref{sec:Results} shows the results in terms of validation, performance, and profiling; Section~\ref{sec:Concl_Future} concludes the paper and presents the future works.

\section{The pipeline}
\label{sec:Pipeline}

The scientific data processing chain, running on a HPC cluster, of modern pipelines for low frequency radio telescopes (aperture arrays) usually includes a signal combination stage, such as, beamforming (e.g.,~\cite{Magro_2015}), where signals from antennas are combined together to form multiple beams directed to different regions of the sky thus harnessing the instantaneous FoV of the radio telescope, essential for radio transients surveys, and/or a correlation engine, which consists of cross-multiplying the signals of every non-redundant baseline of the array to build a sky image (e.g.,~\cite{Clark_Correlation_2013}). 
The output from the beamformer and correlator becomes the input of scientific pipelines having several targets: for example, beamformer's output can be given in input to FRB detection pipelines (e.g., Heimdall), and correlator's output can be given in input to imaging pipelines (e.g., WSClean~\cite{Offringa_WSClean_2014}, CASA~\cite{Collier_CASA_2021}, DDFacet~\cite{Monnier_DDFacet_2022}, and RICK~\cite{De_Rubeis_RICK_2025,Lacopo_RICK_2025}). Alternatively, beamforming and correlation operations can be performed on board FPGA boards~\cite{Peng_Beamformer_FPGA_2024,Morrison_Correlator_FPGA_2023,van_der_Byl_2021}.

These same operations are implemented in the Northern Cross radio telescope processing pipeline~\cite{De_Barro_2025}, which executes the following steps on GPU of an HPC cluster~\cite{Naldi_Pipeline_2025}: (1) the beamformer/correlator; (2) the FRB detection pipeline (i.e. Heimdall); (3) a Machine-Learning (ML)-based classifier to identify true FRB candidates (in our case, FETCH~\cite{Agarwal_FETCH_2020}). The beamformer can run multi-GPU and produce multiple beams, whereas, Heimdall runs on a single GPU and can only process a single beam per time. Heimdall runs on a Docker container to foster portability and reproducibility.

Heimdall is made of a data ingestion section, running on CPU, where data are received either from a SIGPROC filterbank file or a PSRDADA ring buffer. Then, the function \texttt{hd\_execute}, representing the processing core of Heimdall, is called. This function basically coincides with a loop processing \texttt{nsamps\_gulp} time samples per iteration. Before the loop starts, data are copied once from the host to the device. At every iteration, a sequence of different GPU kernels having specific scientific tasks is executed (Chapter 4 of~\cite{Barsdell_PhD_Thesis_2012}). At the end of each iteration, a device-to-host copy involving few bytes is performed, only transferring the metadata, such as time of arrival (TOA) of the signal, dispersion measure (DM), and signal-to-noise ratio (SNR), of all possible FRB candidates ($\sim$$O(10^3)-O(10^4)$) that are saved in the final output files of Heimdall with \texttt{.cand} format. These final \texttt{.cand} files are then passed, after transformation in \texttt{.hdf5} format, to FETCH classifier.

DM is one of the most important quantities to characterize a FRB. When a radio signal propagates from a source to the Earth, it interacts with free electrons in the interstellar medium, and is delayed differently frequency by frequency, with the delay depending on the DM along the line of sight, calculated as~\cite{Barsdell_2012}:
\begin{equation}
    \label{eq:DM}
    {\rm DM} \equiv \int_0^d n_{\rm e} {\rm d}l,
\end{equation}
where $n_{\rm e}$ is the electron number density in~cm$^{-3}$, $d$ is the distance to the source in~pc, thus the DM is expressed in~pc~cm$^{-3}$. Dedispersion is the computational process to correct these frequency-dependent time delays. Since both $n_{\rm e}$ and $d$ are generally not known, DM is guessed over a DM trial grid (dedispersion transform), making dedispersion a highly numerical intensive problem over a 3D grid of \texttt{nsamps\_gulp} time samples per iteration, $N_\nu$ frequency channels, and $N_{\rm DM}$ DM trials. For the Northern Cross, time resolution is $\Delta t_{\rm sample} = 138.24$~$\mu$s and, for the considered data, $N_\nu = 1024$ and frequency resolution is $|\Delta_\nu| = 0.014468$~MHz. A typical value for the number of DM trials is $N_{\rm DM} \sim O(10^3)$.

Dedispersion kernel in Heimdall is GPU-parallelized with the customized and CUDA-based \texttt{dedisp} library\footnote{\url{https://github.com/ajameson/dedisp}; Jameson \& Barsdell}, whereas all the other GPU modules are parallelized with Thrust C++ Template Library~\cite{Hoberock_and_Bell_Thrust_2010}, supplied as part of NVIDIA CUDA toolkit\footnote{\url{http://developer.nvidia.com/cuda-downloads}}.

\subsection{The offline or SIGPROC mode}
\label{sec:Pipeline_Offline}

When talking about offline (or SIGPROC) or real-time (or PSRDADA) modes of the pipeline, we refer to the entire pipeline including the beamformer and Heimdall working together. In the offline mode, the beamformer writes its output in a SIGPROC filterbank file, which Heimdall reads in input. SIGPROC has been used by numerous pulsar astronomers since 2001 (e.g.,~\cite{Camilo_SIGPROC_2002,Breton_SIGPROC_2008,Rickett_SIGPROC_2014,Pan_SIGPROC_2016,Kirsten_SIGPROC_2021,Turner_SIGPROC_2025}). 

A SIGPROC file is structured with a header of metadata followed by a stream of data bytes. The header contains information about the data acquisition, such as the source name, the observation length (i.e., the acquisition time), the frequency of the first channel ($\nu_0$), the bandwidth of a single channel or frequency resolution ($\Delta_\nu$), the number of frequency channels ($N_\nu$), the number of bits per time sample in each frequency channel ($N_{\rm bits}$), and the time sample resolution ($\Delta t_{\rm sample}$). In our case, we have $N_\nu$, $\Delta t_{\rm sample}$, and $\Delta_\nu$ reported at the end of Section~\ref{sec:Pipeline}, with $\Delta_\nu$ taken with minus sign, $\nu_0 = 415.854456$~MHz, and $N_{\rm bits} = 16$, whereas the other parameters depend on the specific observation. The frequency of the last channel is $\nu_{1023} = \nu_0 + (N_\nu - 1) \times \Delta_\nu = 401.053992$~MHz.
The stream of data bytes is memorized as a sequence of spectra in time, where the values $I(t_i,\nu_j)$ of the signal intensity, typically expressed in ADC counts proportional to the received power, are stored for every time sample $t_i$ and frequency channel $\nu_j$, in a time-major, frequency-minor order. Having $N_{\rm bits} = 16$~bits, every single spectrum occupies:
\begin{equation}
    \label{eq:Mem_Spectrum}
    Mem_{\rm spectrum} = N_\nu \times \frac{N_{\rm bits}}{8} = 2048\text{ bytes.} 
\end{equation}

After the end of beamformer's execution, Heimdall is sequentially launched with the following command:
\begin{lstlisting}[language=bash, label={lst:Heimdall_sigproc_generical}]
heimdall -f SIGPROC_file.fil
\end{lstlisting}
where \texttt{SIGPROC\_file.fil} is the path and name of the input SIGPROC filterbank file. 
In a single-beam configuration, one GPU node is sufficient, to execute the beamformer and Heimdall processes in sequence.

\subsection{The real-time or PSRDADA mode}
\label{sec:Pipeline_Real-Time}

A PSRDADA ring buffer has a fixed memory area organized in \texttt{<n>} blocks each of \texttt{<b>} bytes size, and needs a pointer writing new data (producer) and a pointer reading the data (consumer). When the end of the ring buffer is reached, data writing wraps around to the beginning. If the buffer is completely full, the producer cannot write new data blocks until the consumer reads existing data to free up space. Data is structured within the PSRDADA ring buffer using a layout similar to that of a SIGPROC file, i.e., with a metadata header followed by a stream of data bytes, although the parameters in the metadata header are differently parsed as in the SIGPROC header (see Section~\ref{sec:Tests_Writer}).

In the real-time mode, prior to the beamformer-Heimdall execution, the PSRDADA ring buffer has to be created with the following command:
\begin{lstlisting}[language=bash, label={lst:dada_creation}]
dada_db -k dada -b <b> -n <n>
\end{lstlisting}
where \texttt{dada} is an identifier key. This command can be directly launched from command line or through a script or a process manager (see Section~\ref{sec:Concl_Future}).  

After ring buffer creation, differently from the offline mode, the first process to be launched is Heimdall, which represents the consumer process. Heimdall is launched with command:
\begin{lstlisting}[language=bash, label={lst:Heimdall_psrdada_generical}]
heimdall -k dada
\end{lstlisting}
where \texttt{dada} has to be the same dada key of the previously created PSRDADA ring buffer. Heimdall process waits to start its execution until the producer process is launched.

The second process to be launched is the beamformer, representing the producer process. The execution of the beamformer and Heimdall is asynchronous: while data are produced by the beamformer and written to the ring buffer, Heimdall reads, i.e., ``consumes'', the data from the ring buffer and processes them at chunks of \texttt{nsamp\_gulp} elements. The execution times of Heimdall alone launched in SIGPROC and PSRDADA modes are comparable, since GPU processing, representing the computational core of Heimdall, is the same and the only difference is the data ingestion mechanism. The real advantage of PSRDADA over SIGPROC mode is the asynchronous rather than sequential execution of the producer and consumer processes. If the producer is faster than the consumer, a waiting queue occurs until new memory regions of the ring buffer are available when data are read by the consumer. If the consumer is faster than the producer, the ring buffer stays in an ``idle'' state, until new data are written to the ring buffer by the producer. In an ideal case, the consumer and producer processes have to proceed with a comparable speed to avoid waiting times and keep up with the real-time regime (Sections~\ref{sec:Results_Performance_Default_Parameters} and~\ref{sec:Results_Performance_Varying_Parameters}). In a single-beam configuration, we need a GPU node for the beamformer and a GPU node for Heimdall, whose resources have to be simultaneously required.

\section{Testbed setup}
\label{sec:Tests}

\subsection{The writer}
\label{sec:Tests_Writer}
The beamformer-Heimdall pipeline in real-time is still a work in progress (Section~\ref{sec:Concl_Future}). To validate the real-time mode against the offline mode, we expressly build a test pipeline, where the beamformer is replaced by a CPU-based writer, taking in input the data of a SIGPROC filterbank file, so that the two modes have the same input data. Heimdall runs on the GPU of the same node where the writer runs on the CPU. The writer emulates the data transmission interface between the beamformer and Heimdall, maintaining the same pipeline's structure, and is organized as follows: (1) {\bf Reading phase.} The writer reads the stream of data bytes with \texttt{fread} function after the header location of a SIGPROC filterbank file, storing it in a buffer in memory; (2) {\bf PSRDADA header preparation.} PSRDADA header cannot be directly read from SIGPROC header but has to be prepared with a specific function, since metadata parameters are differently parsed in the two modes. PSRDADA header contains \texttt{FREQ} that, differently from SIGPROC header, is not the frequency of the first channel $\nu_0$ but the central frequency of the bandwidth, $FREQ = 0.5 \times (\nu_0 + \nu_{1023})$, and $BW = \Delta_\nu \times N_\nu = -14.815232$~MHz, the total frequency bandwidth taken with the right sign. Using $FREQ = \nu_0$, as in SIGPROC header, and $BW$ in absolute value would lead to frequencies shifted of $-|BW|/2$, resulting in a DM underestimation of $\sim$$5\%$, following Eq. (1) in~\cite{Barsdell_2012}. This bug was encountered during tests and successively fixed; (3) {\bf Writing phase.} A while loop writes the buffer in the PSRDADA ring buffer in chunks of \texttt{CHUNK\_SIZE} bytes per iteration, through \texttt{ipcio\_write} function of PSRDADA framework.
Because the structure of the writer-Heimdall test pipeline is identical to that of the future beamformer-Heimdall pipeline, its analysis provides valuable insights into the final configuration. These insights will be crucial when the system is deployed to production for the Northern Cross, and can be extrapolated in favour of other radio telescopes to meet stringent timing requirements.

The writer is compiled as:
\begin{lstlisting}[language=bash]
gcc writer_psrdada.c -o writer_psrdada -I/<PSRDADA_INSTALLATION_PATH>/include -L/<PSRDADA_INSTALLATION_PATH>/lib -lpsrdada
\end{lstlisting}


\subsection{The launching commands sequence}
\label{sec:Tests_Launching_Commands}
The offline mode only needs to run a single command:
\begin{lstlisting}[language=bash, label={lst:Heimdall_sigproc}]
heimdall -f SIGPROC_file.fil -output_dir ./cand_SIGPROC -dm 10 3000 -dm_tol 1.001 -boxcar_max 512 -gpu_id 0 -v > output_SIGPROC.txt
\end{lstlisting}
where: (1) \texttt{cand\_SIGPROC} is the output directory where the \texttt{.cand} files are saved; (2) we explore a large 
DM range of $[10,3000]$~pc~cm$^{-3}$ (see Section~\ref{sec:Pipeline}) with a proper DM resolution of \texttt{dm\_tol} $= 1.001$~pc~cm$^{-3}$, for FRB blind searches; (3) \texttt{-boxcar\_max 512} means that Heimdall looks for FRB signal bursts large up to $512 = 2^9$ time samples, corresponding to a temporal width of $2^9 \times 138.24$~$\mu$s = $70.8$~ms, where $138.24$~$\mu$s is $\Delta t_{\rm sample}$ (Section~\ref{sec:Pipeline}). This is a reasonable upper limit for bursts width search (e.g.,~\cite{Wei_FRB_Burst_Duration_2025}). 
To avoid edge effects in the signal search, Heimdall only considers $\sim$\texttt{nsamps\_gulp - boxcar\_max} samples per chunk of the \texttt{hd\_execute} function loop as valid samples; (4) \texttt{-gpu\_id 0} specifies that Heimdall runs on the GPU with ID 0; (5) \texttt{-v} enables output's verbosity. 

The real-time mode requires more launching commands (Section~\ref{sec:Pipeline_Real-Time}). 
To be sure that no PSRDADA processes with \texttt{dada} key are pending, 
a PSRDADA ring buffer deletion command is executed:
\begin{lstlisting}[language=bash, label={lst:dada_deletion}]
dada_db -k dada -d
\end{lstlisting}
Secondly, the PSRDADA ring buffer is created, adopting \texttt{<b>} = 128 MiB and \texttt{<n>} = 8 as default parameter combination (Section~\ref{sec:Tests_Parameters}), for a total ring buffer size of 1~GiB.
Thirdly, Heimdall (the consumer) is launched:
\begin{lstlisting}[language=bash, label={lst:Heimdall_dada}]
heimdall -k dada -output_dir ./cand_PSRDADA -dm 10 3000 -dm_tol 1.001 -boxcar_max 512 -gpu_id 0 -v > output_PSRDADA.txt &
\end{lstlisting}
with the same runtime parameters as Heimdall in the offline mode except for \texttt{-k dada}. 
The \texttt{dada} key has to be the same in PSRDADA deletion, creation, and Heimdall's commands.
At last, the writer (the producer) is launched:
\begin{lstlisting}[language=bash, label={lst:dada_writer}]
./writer_psrdada SIGPROC_file.fil > output_writer_psrdada.txt
\end{lstlisting}
where \texttt{SIGPROC\_file.fil} is the path and name of the input SIGPROC filterbank file.

\subsection{The input files}
\label{sec:Tests_Files}
To validate the real-time mode against the offline mode of the pipeline, and to compare their performances, we employed two classes of reference SIGPROC filterbank files, containing real data acquired by 16 cylinders of the NS arm of the Northern Cross radio telescope (Section~\ref{sec:Intro}):
\begin{enumerate}
    \item 11 files with no FRB detections where 12 synthetic FRBs, with different values of TOA, DM, and Signal-to-Noise Ratio (SNR), were injected in each of them with FRB Faker software\footnote{\url{https://gitlab.com/houben.ljm/frb-faker}} (hereafter, called ``synthetic files''). Some of the injected FRBs have a DM larger than 3000~pc~cm$^{-3}$: for these files, we adopt an upper limit larger than 3000~pc~cm$^{-3}$ for DM range in Heimdall commands (Section~\ref{sec:Tests_Launching_Commands}).
    \item 11 files containing one true FRB detection each, having known TOA and DM (hereafter, called ``true files'').
\end{enumerate}
Having reference values, we do not run the classifier after Heimdall's execution but we look for the corresponding FRB candidates in the \texttt{.cand} files.

\subsection{The parameter space}
\label{sec:Tests_Parameters}

We run the offline and real-time executions for each of the files with this default combination of parameters: \texttt{nsamps\_gulp} of $262144 = 2^{18}$ (default Heimdall parameter, common to offline and real-time modes), \texttt{<b>} $= 128$~MiB, \texttt{<n>} $= 8$, and \texttt{CHUNK\_SIZE} $= 8$~MiB (only for real-time mode).
To make the validation process more robust, we select synthetic file n. 1 and true file n. 10, and we repeat the search of the FRBs exploring the parameter space according to Table~\ref{tab:Parameters}.
\begin{table}[!htp]\centering
\caption{Parameter space explored in the tests.}\label{tab:Parameters}
\begin{tabular}{lcccc}
\toprule
{\bf Combination ID} & \texttt{nsamps\_gulp} &\texttt{<b>} & \texttt{<n>} &\texttt{CHUNK\_SIZE} \\
 & & [MiB] & & [MiB] \\
 \midrule
(1) & $2^{18}$ & 128 & 8 & 8\\
(2) & $2^{19}$ & 128 & 8 & 8\\
(3) & $2^{20}$ & 128 & 8 & 8\\
(4) & $2^{21}$ & 128 & 8 & 8\\
(5) & $2^{21}$ & 256 & 4 & 8\\
(6) & $2^{21}$ & 512 & 2 & 8\\
(7) & $2^{18}$ & 512 & 2 & 512\\
(8) & $2^{21}$ & 4096 & 2 & 4096\\
\bottomrule
\end{tabular}
\end{table}
The first line (combination (1)) corresponds to the default parameter combination. 
In cases (1)-(4), explored for both modes, we only increase \texttt{nsamps\_gulp} by powers of two, while keeping the PSRDADA parameters fixed. To set a different \texttt{nsamps\_gulp}, the further \texttt{-nsamps\_gulp <nsamps\_gulp>} runtime parameter has to be passed to Heimdall launching commands in Section~\ref{sec:Tests_Launching_Commands}.  
In cases (4)-(6), we explore three different configurations of \texttt{<b>} and \texttt{<n>} with the highest \texttt{nsamps\_gulp}, maintaining the total PSRDADA ring buffer size at $1$~GiB and \texttt{CHUNK\_SIZE} $= 8$~MiB. In cases (7)-(8), we explore two configurations when \texttt{nsamps\_gulp}, \texttt{<b>}, and \texttt{CHUNK\_SIZE} occupy the same number of bytes. 
The number of processed bytes corresponding to \texttt{nsamps\_gulp} is calculated as:
\begin{equation}
    \label{eq:Mem_nsamps_gulp}
    Mem_{\rm nsamps\_gulp} = {\rm nsamps\_gulp} \times Mem_{\rm spectrum},
\end{equation}
where $Mem_{\rm spectrum}$ is given by Eq.~\eqref{eq:Mem_Spectrum}.
For \texttt{nsamps\_gulp} $= 2^{18}$ and $2^{21}$, $Mem_{\rm nsamps\_gulp}$ is $512$~MiB and $4096$~MiB, respectively. In case (8), the size of the PSRDADA ring buffer is $8$~GiB, 8x compared to other cases. 

\subsection{The employed hardware}
\label{sec:Hardware}

We run our test on a single GPU of \texttt{med-node001} node based in Medicina, with the following hardware features:
\begin{enumerate}
    \item CPU: Dual-socket Intel Xeon Gold 5416S @ 2.0 GHz (Turbo up to 4.0 GHz), 32 cores / 64 threads total (16 cores per socket, Hyper-Threading enabled);
    \item RAM Memory: 2 TiB;
    \item GPU: 2 x 46068 MiB VRAM GPU NVIDIA L40S. 92136 MiB total GPU memory of the node;
\end{enumerate}

\section{Results}
\label{sec:Results}

\subsection{PSRDADA vs SIGPROC validation}
\label{sec:Results_Validation}

\subsubsection{Default parameters}
\label{sec:Results_Validation_Default_Parameters}

\begin{figure}[ht]
\centering
\subfloat[TOA results.]{\includegraphics[width=0.48\textwidth]{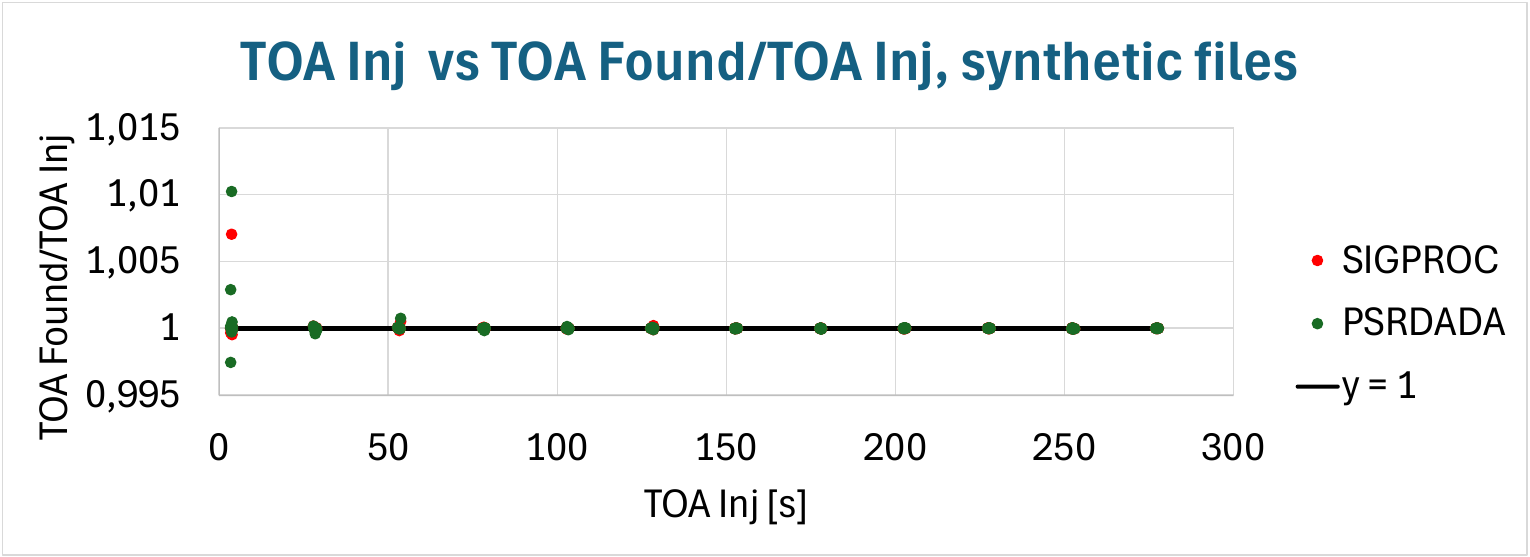} \label{fig:Synthetic_Default_Validation_TOA}}
\hfill
\subfloat[DM results.]{\includegraphics[width=0.48\textwidth]{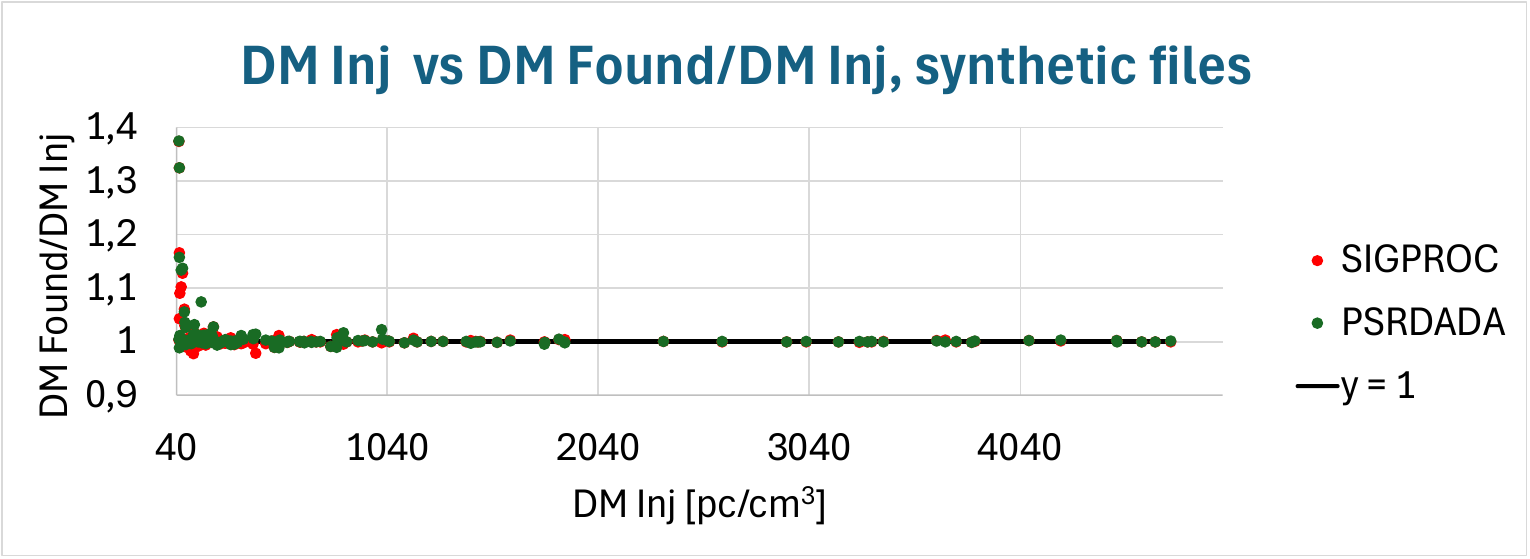}  \label{fig:Synthetic_Default_Validation_DM}}
\hfill
\subfloat[SNR results.]{\includegraphics[width=0.48\textwidth]{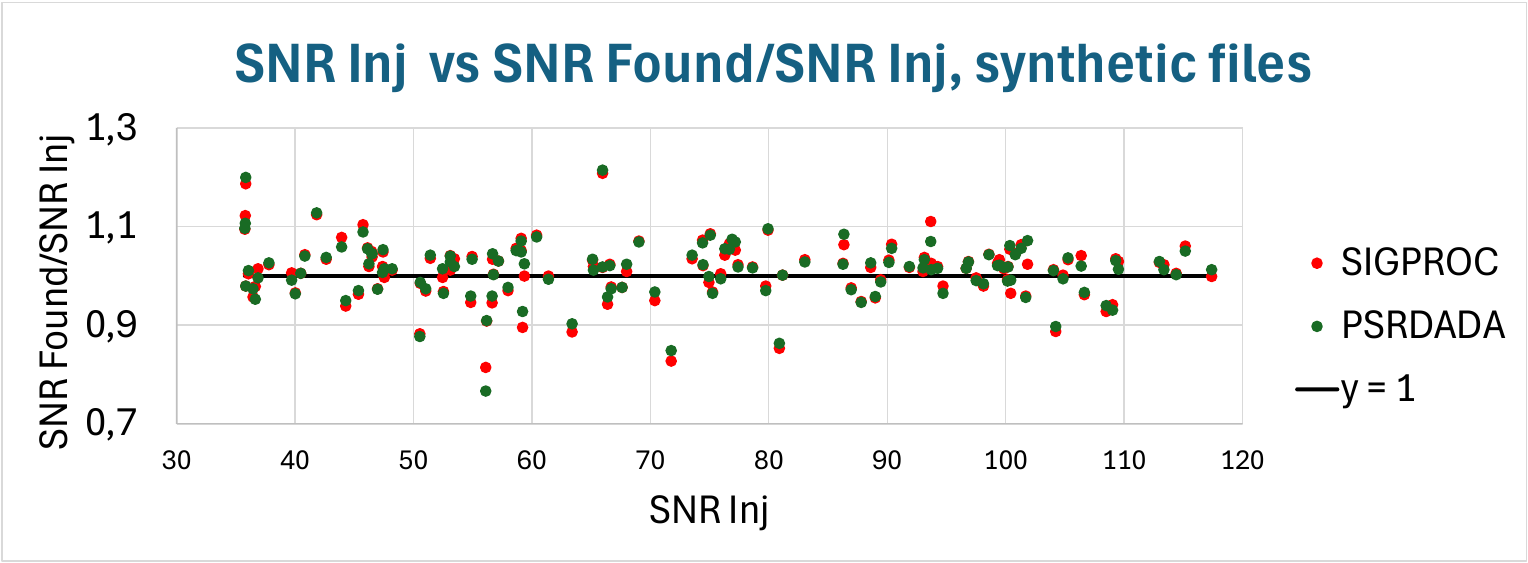} \label{fig:Synthetic_Default_Validation_SNR}}
\hfill
\caption{Ratios between the obtained FRB parameters and injected FRB parameters for the 11 synthetic files with the default parameter combination in Table~\ref{tab:Parameters}.}
\label{fig:Synthetic_Default_Validation}
\end{figure}

\begin{figure}[ht]
\centering
\subfloat[TOA results.]{\includegraphics[width=0.48\textwidth]{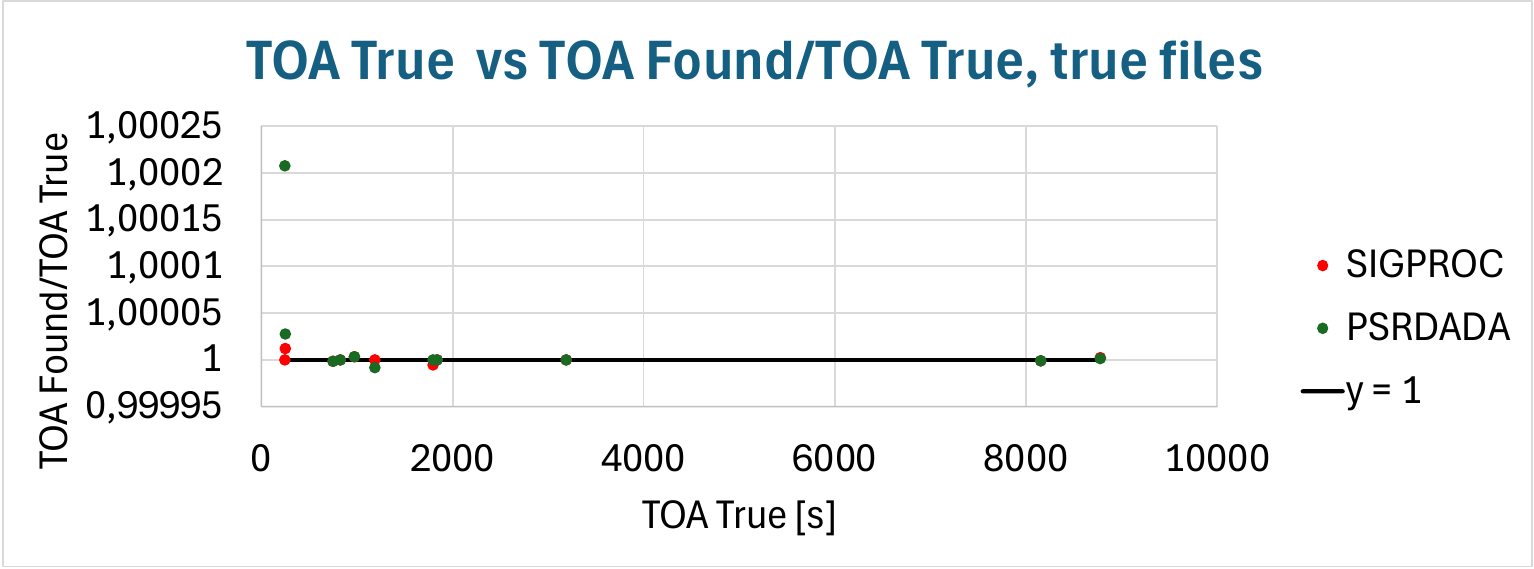} \label{fig:True_Default_Validation_TOA}}
\hfill
\subfloat[DM results.]{\includegraphics[width=0.48\textwidth]{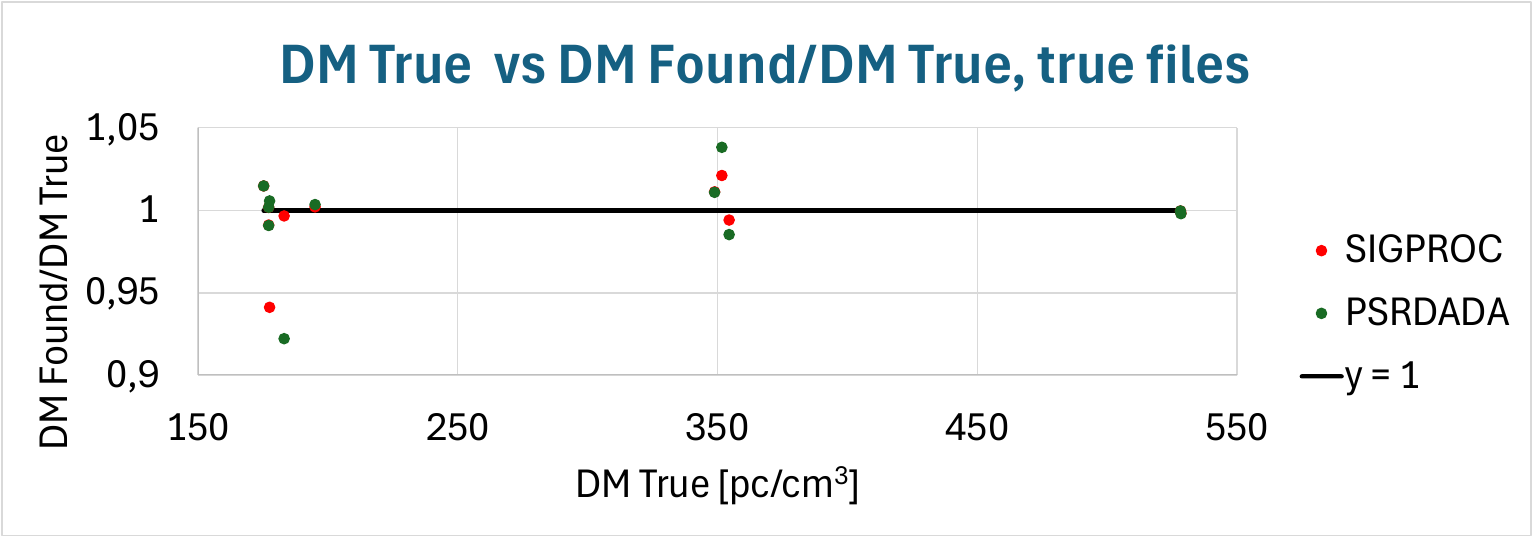}  \label{fig:True_Default_Validation_DM}}
\hfill
\caption{Ratios between the obtained FRB parameters and true FRB parameters for the 11 true files with the default parameter combination in Table~\ref{tab:Parameters}.}
\label{fig:True_Default_Validation}
\end{figure}

Figures~\ref{fig:Synthetic_Default_Validation} and~\ref{fig:True_Default_Validation} illustrate the reference FRB parameters on the x-axis and the ratios between the found FRB parameters and the reference FRB parameters on the y-axis, for synthetic and true files, respectively, adopting the default parameter combination (line (1) in Table~\ref{tab:Parameters}). SIGPROC and PSRDADA ratios are represented in red and green, respectively. 
The $y = 1$ line is shown as a reference. Figures~\ref{fig:Synthetic_Default_Validation_TOA},~\ref{fig:Synthetic_Default_Validation_DM}, and~\ref{fig:Synthetic_Default_Validation_SNR} report results for TOA, DM, and SNR, for all synthetic files, whereas Figures~\ref{fig:True_Default_Validation_TOA} and~\ref{fig:True_Default_Validation_DM} report results for TOA and DM, for all true files. 

The figures display a very good agreement of the obtained FRB parameters with the reference values, specifically for TOA and DM, where the ratios narrowly distribute around 1. All the TOA ratios, for SIGPROC and PSRDADA, for synthetic and true files, are between $0.97$ and $1.03$. For the synthetic files, $99.24\%$ of the TOA points distribute in the narrower range $[0.995,1.005]$, for both modes; for the true files, these percentages rise to $100\%$. For DM ratios, $93.13\%$ points remain in the $[0.97,1.03]$ range for both modes, for synthetic files, and the same range is occupied by $90.91\%$ (SIGPROC) and $81.82\%$ (PSRDADA), for true files. When considering the $[0.995,1.005]$ range, the percentages are: $74.05\%$ (SIGPROC, synthetic), $74.81\%$ (PSRDADA, synthetic), $45.45$\% (SIGPROC, true), and $36.36\%$ (PSRDADA, true). It is important to highlight the minor statistics for the true files, containing one FRB each. SNR ratios from synthetic files present a larger distribution, due to the fact that the FRB injection software (FRB Faker, Section~\ref{sec:Tests_Files}) and the detection software (Heimdall) follow different procedures for noise and signal power estimation. To avoid discrepancies between injected and detected SNRs, a common estimation method should be established for both the injection and detection procedures. Nevertheless, SNR ratios still settle in a reasonable range: $43.41\%$ (SIGPROC) and $48.09\%$ (PSRDADA) in the $[0.97,1.03]$ range, and $90.08\%$ (SIGPROC) and $93.13\%$ (PSRDADA) in the $[0.90,1.10]$ range.

\subsubsection{Varying parameters}
\label{sec:Results_Validation_Varying_Parameters}

Exploring the parameter space of Table~\ref{tab:Parameters} does not affect the results accuracy: with true file n. 10, the TOA ratios of the single FRB across the different parameter combinations are always exactly equal to 1 and the DM ratios  span between $0.997$ and $1.003$, for both modes. 
With synthetic file n. 1, $100\%$ of TOA ratios of the 12 FRBs obtained with combinations (1)-(4) in SIGPROC case are in the $[0.995,1.005]$ range, $91.11\%$ of DM ratios are in the $[0.97,1.03]$ range, and $91.11\%$ of SNR ratios are in the $[0.9,1.1]$ range. With combinations (1)-(8) in PSRDADA case, the percentages in the same ranges for TOA, DM, and SNR are: $100\%$, $91.01\%$, and $91.01\%$. These percentages also reflect those obtained for the 12 FRBs within each parameter combination.

\subsection{Performance considerations}
\label{sec:Results_Performance}

With this test pipeline, performance assessment is still preliminary but already provides some important considerations in perspective of the future pipeline. In the real-time pipeline, the writer emulates the beamformer. Its presence is essential since, in a PSRDADA ring buffer data structure, we need a writing pointer that produces the data (i.e., the writer) and a reading pointer that consumes the data (i.e., Heimdall). For a fair performance comparison between offline and real-time modes of the pipeline, we need to define an analog of the beamformer in the offline mode as well. We, therefore, build a ``SIGPROC writer'' that: (1) reads the SIGPROC filterbank stream of data bytes after its header location in a buffer, as in the PSRDADA writer; (2) prepares the SIGPROC header and writes it in a new SIGPROC filterbank file; (3) writes the content of the buffer in the new SIGPROC filterbank file, after the header location. The total time of the offline pipeline is, therefore: 
\begin{equation}
    \label{eq:time_offline_pipeline}
    t_{\rm offline} = t_{\rm writer, SIGPROC} + t_{\rm Heimdall, SIGPROC},
\end{equation}
where $t_{\rm writer, SIGPROC}$ and $t_{\rm Heimdall, SIGPROC}$ are the execution times of SIGPROC writer and Heimdall launched in offline mode (Section~\ref{sec:Tests_Launching_Commands}).

Instead, the total time of the real-time pipeline is:
\begin{equation}
    \label{eq:time_real-time_pipeline}
    t_{\rm real-time} = {\rm Max}(t_{\rm writer, PSRDADA},  t_{\rm Heimdall,PSRDADA}),
\end{equation}
since the two processes are asynchronous, where $t_{\rm writer, PSRDADA}$ and $t_{\rm Heimdall,PSRDADA}$ are the execution times of PSRDADA writer and Heimdall launched in real-time mode (Section~\ref{sec:Tests_Launching_Commands}).

\subsubsection{Default parameters}
\label{sec:Results_Performance_Default_Parameters}

\begin{figure*}[ht]
\centering
\subfloat[Execution times against acquisition times for all the synthetic files.]{\includegraphics[width=0.4\textwidth]{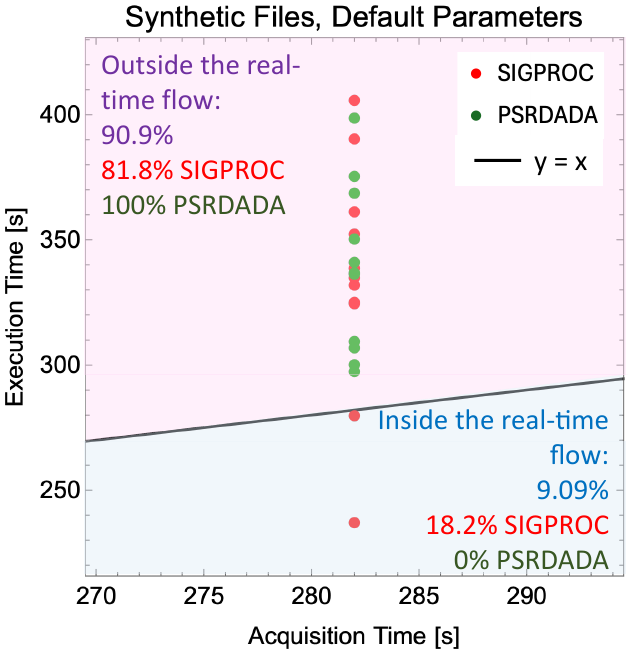} \label{fig:Default_Acq_Times_vs_Exec_Times_Synthetic}}
\hfill
\subfloat[Execution times against acquisition times for all the true files.]{\includegraphics[width=0.42\textwidth]{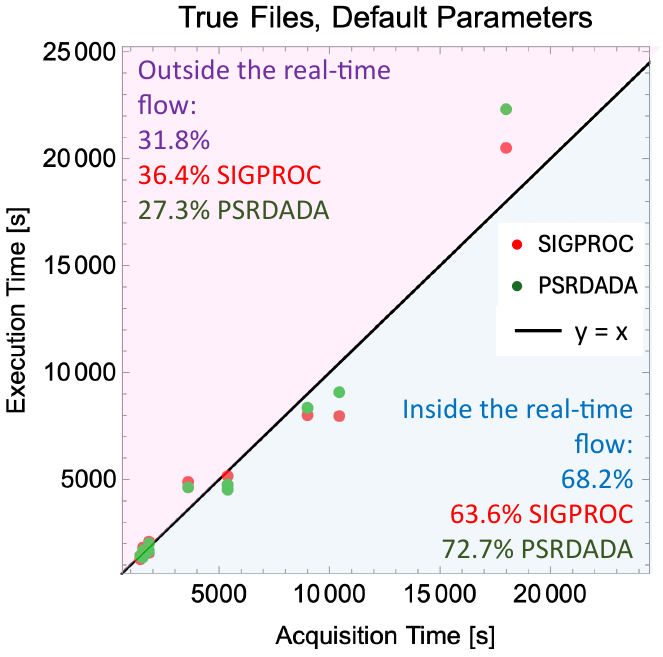}  \label{fig:Default_Acq_Times_vs_Exec_Times_True}}
\hfill
\subfloat[PSRDADA times against SIGPROC times for all the synthetic files.]{\includegraphics[width=0.4\textwidth]{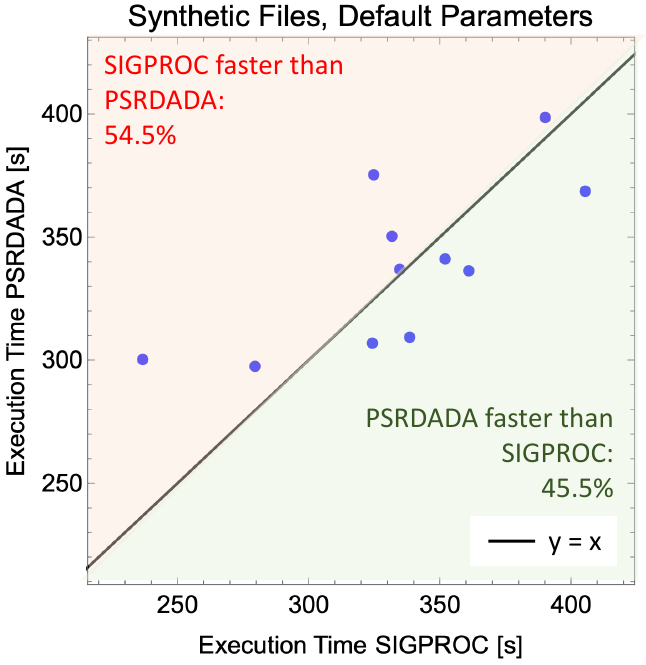} \label{fig:Default_SIGPROC_Times_vs_PSRDADA_Times_Synthetic}}
\hfill
\subfloat[PSRDADA times against SIGPROC times for all the true files.]{\includegraphics[width=0.42\textwidth]{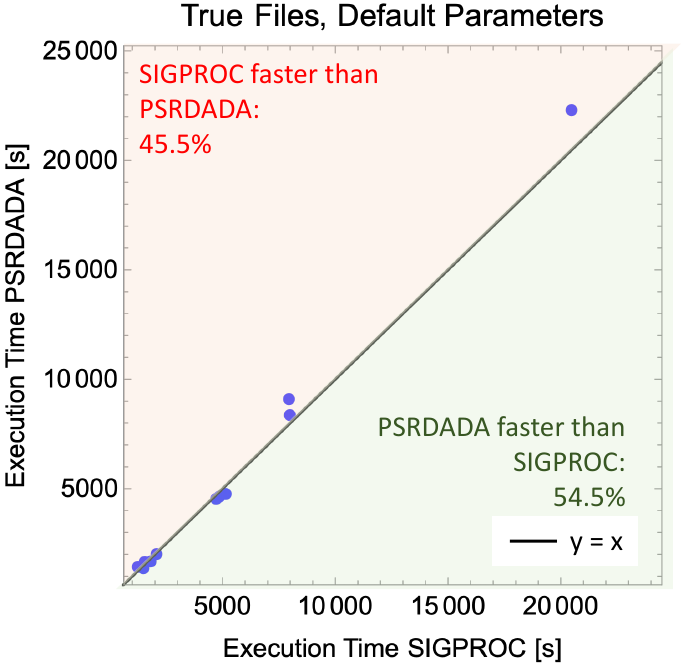}  \label{fig:Default_SIGPROC_Times_vs_PSRDADA_Times_True}}
\hfill
\caption{Execution times against acquisition times for the default parameter combination (Figures~\ref{fig:Default_Acq_Times_vs_Exec_Times_Synthetic} and~\ref{fig:Default_Acq_Times_vs_Exec_Times_True}), where SIGPROC and PSRDADA points are in red and green, respectively. PSRDADA times against SIGPROC times for the default parameter combination (Figures~\ref{fig:Default_SIGPROC_Times_vs_PSRDADA_Times_Synthetic} and~\ref{fig:Default_SIGPROC_Times_vs_PSRDADA_Times_True}).}
\label{fig:Default_Times}
\end{figure*}

Figure~\ref{fig:Default_Times} shows the performance results with the default parameter combination. Figures~\ref{fig:Default_Acq_Times_vs_Exec_Times_Synthetic} and~\ref{fig:Default_Acq_Times_vs_Exec_Times_True} report the execution times obtained with the 11 synthetic files and with the 11 true files, respectively, against the data acquisition times. SIGPROC and PSRDADA points are red and green, respectively. The one-to-one relation, or ``real-time line'', is represented as a black solid line and divides the plane in two distinct regions, the ``real-time'' region (light blue) and the ``outside real-time'' region (light purple).

All synthetic files have an acquisition time of $4.7$~min. Their execution times span the $[3.9,6.8]$~min range. Figure~\ref{fig:Default_Acq_Times_vs_Exec_Times_Synthetic} shows that all the SIGPROC points except for two points and all the PSRDADA points fall outside the real-time region. This can be due to the fact that, with such small involved times, the pipeline's overheads (e.g., startup operations, such as, CUDA initialization and memory allocations and shutdown operations, such as, writing and closing of all \texttt{.cand} files and memory freeing) and synchronization's operations (e.g., connection to ring buffer in PSRDADA case) can have a not negligible weight in the total time of the pipeline, which results to be no longer dominated by GPU processing. A more realistic scenario is provided by the true files. We can see that the points are more closely distributed around the real-time line and that the majority of the points ($63.6\%$ for SIGPROC) and ($72.7\%$ for PSRDADA) meet real-time contraints. With execution times ranging from $20.7$ minutes to $6.2$ hours, GPU processing once again dominates.

Figures~\ref{fig:Default_SIGPROC_Times_vs_PSRDADA_Times_Synthetic} and~\ref{fig:Default_SIGPROC_Times_vs_PSRDADA_Times_True} display the PSRDADA execution times against the SIGPROC execution times for synthetic and true files, respectively, with default parameters. 
The one-to-one relation is again shown as a reference, to divide the plane in the region where PSRDADA is faster than SIGPROC (light green) and, conversely, where SIGPROC is faster than PSRDADA (light red). In both cases, the points are distributed around the one-to-one relation and are $\sim$50\% and $\sim$50\% divided between the two regions, more scattered for the synthetic files.  
The similarity between $t_{\rm real-time}$ and $t_{\rm offline}$ can be explained by the fact that $t_{\rm Heimdall, SIGPROC} \sim t_{\rm Heimdall, PSRDADA}$, since GPU processing is the same in the two Heimdall's modes, and $t_{\rm writer, SIGPROC} \sim 1-2\%\text{ }t_{\rm Heimdall, SIGPROC}$. This implies that asynchronism between PSRDADA writer and PSRDADA Heimdall does not provide a significant advantage over SIGPROC writer and SIGPROC Heimdall executed in sequence. A different scenario is expected when the two writers are replaced by the beamformer, as its execution time will be significantly longer than that of the SIGPROC writer.

\subsubsection{Varying parameters}
\label{sec:Results_Performance_Varying_Parameters}

\begin{figure*}[ht]
\centering
\subfloat[Speedup for synthetic file n. 1. The Speedup $= 1.18$x (dashed red) and $= 1.24$x (dashed green) lines are shown as a reference, to identify the required speedups that SIGPROC and PSRDADA modes should have to enter in the real-time flow.]{\includegraphics[width=0.8\textwidth]{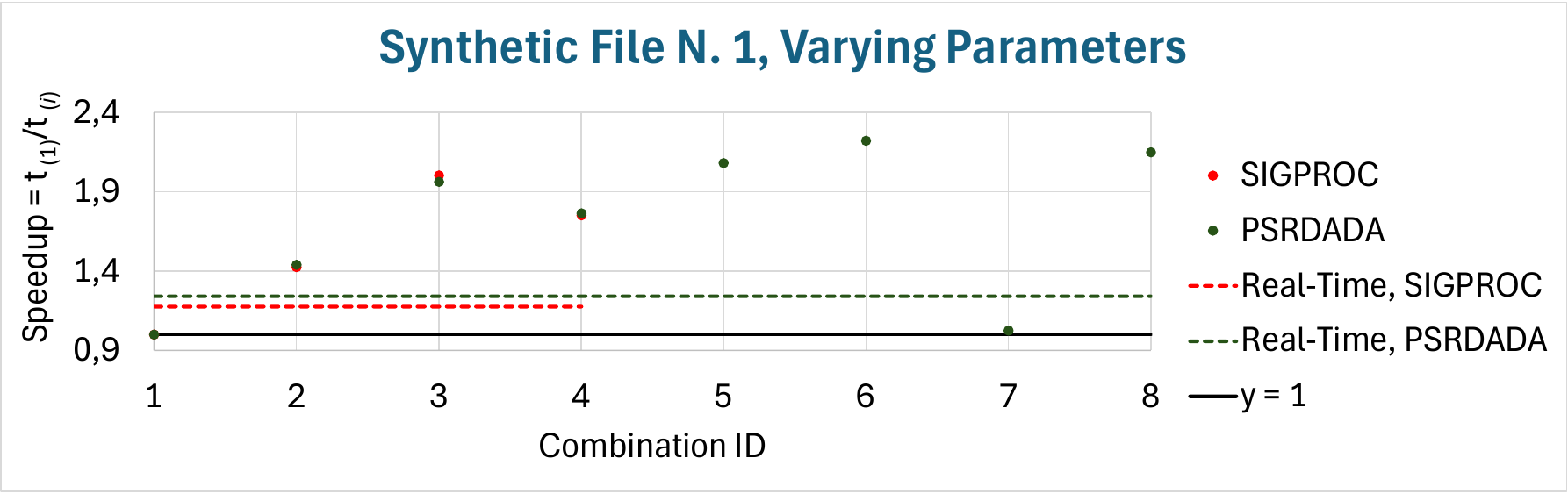} \label{fig:Varying_Parameters_Speedups_Synthetic}}
\hfill
\subfloat[Speedup for true file n. 10.]{\includegraphics[width=0.8\textwidth]{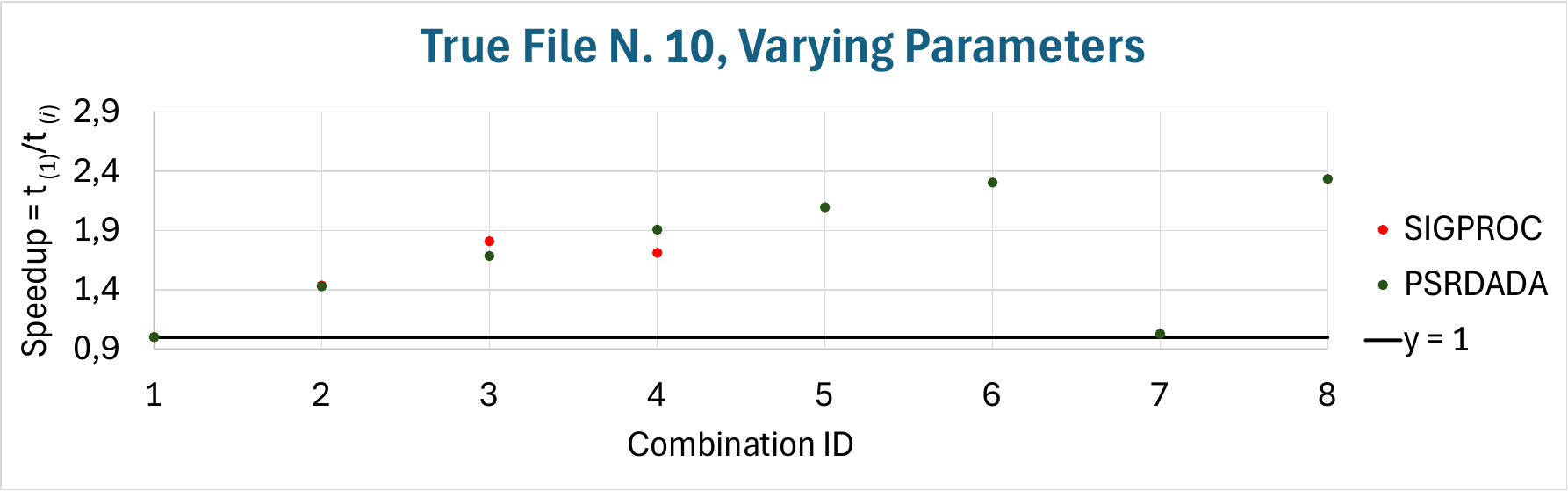}  \label{fig:Varying_Parameters_Speedups_True}}
\hfill
\caption{Speedup against parameter combination ID. Speedup $= 1$x lines are shown as a reference.}
\label{fig:Varying_Parameters_Speedups}
\end{figure*}

Figure~\ref{fig:Varying_Parameters_Speedups} illustrates the Speedup, calculated as the ratio between execution time of combination (1), $t_{(1)}$ and the execution times obtained with each parameter combination, $t_{(i)}$, as a function of the parameter combination ID for synthetic file n. 1 (Figure~\ref{fig:Varying_Parameters_Speedups_Synthetic}) and true file n. 10 (Figure~\ref{fig:Varying_Parameters_Speedups_True}). Figures~\ref{fig:Varying_Parameters_Speedups_Synthetic} and~\ref{fig:Varying_Parameters_Speedups_True} also display the Speedup $= 1$ black solid line as a reference.

Combination (1) (default) provides the worst performance in any case (Speedup = 1x). In combinations (1)-(4), where we consider \texttt{nsamps\_gulp} $= \{1x,2x,4x,8x\}$ and we do not change PSRDADA parameters with respect to default combination, we can see that \texttt{nsamps\_gulp} has a strong impact on the performance. Increasing its value up to $2^{20}$ produces a performance boost for all cases. The increasing trend of the speedup continues up to $2^{21}$ only in the case of the true file with PSRDADA. Doubling \texttt{nsamps\_gulp} does not double the speedup: with $2^{20}$ (4x), we reach a maximum speedup of $2.00$x for SIGPROC with the synthetic file; with $2^{21}$ (8x), we reach a maximum speedup of $1.91$x for PSRDADA with the true file. The parameter \texttt{nsamps\_gulp} determines the size of the time samples chunk processed on the GPU per iteration in the main execution loop of Heimdall. With larger \texttt{nsamps\_gulp}, Heimdall processes fewer gulps, reducing the computational overhead, which can explain the achieved performances. Future works might involve a deeper analysis with \texttt{nvidia\_smi} command and tools such as NVIDIA Nsight Compute profiler\footnote{https://developer.nvidia.com/nsight-compute}, to verify a possible connection between \texttt{nsamps\_gulp} and GPU occupancy, thereby providing further insights into these results.

Concerning PSRDADA, combinations (4), (5), (6), and (8) explore different parameters, while keeping \texttt{nsamps\_gulp} fixed to $2^{21}$. Combinations (4), (5), and (6) maintain the size of the ring buffer constant to 1~GiB, increasing \texttt{<b>} and decreasing \texttt{<n>}, and \texttt{CHUNK\_SIZE} constant to 8~MiB. Increasing \texttt{<b>} and decreasing \texttt{<n>} produces a speedup improvement, passing from $1.76$x to $2.22$x for the synthetic file, and from $1.91$x to $2.30$x for the true file. This can be explained by a mechanism similar to the one observed when increasing \texttt{nsamps\_gulp}: with larger and less blocks, we reduce the consequent overhead. Combinations (7) and (8) have \texttt{nsamps\_gulp}, \texttt{<b>}, and \texttt{CHUNK\_SIZE} occupying the same number of bytes. Combination (7) provides a slight speedup over combination (1) ($1.02$x for both synthetic and true file, respectively), whereas combination (8) provides the highest speedup for the true file ($2.33$x), but its performance is slightly worse than combination (6) for the synthetic file ($2.15$x).

This test provides a significant outcome: properly tuning the involved parameters allows the execution to meet real-time constraints. Figure~\ref{fig:Varying_Parameters_Speedups_Synthetic} also displays the Speedup $= 1.18$x dashed red line and $1.24$x dashed green line, representing the required speedups that SIGPROC and PSRDADA modes should have to enter in the real-time regime for synthetic file n. 1. All parameter combinations, except (1), exceed these speedup factors. Figure~\ref{fig:Varying_Parameters_Speedups_True} does not show the corresponding lines, as the execution times for true file n. 10 meet real-time constraints for every parameter combination.

It is worth noting how the execution times of the two pipeline processes (the writer and Heimdall) vary across the parameter space for both SIGPROC and PSRDADA modes. 
For SIGPROC, the writer is a constant offset, not depending on the adopted parameters ($3.95$~s for the synthetic file and $62.82$~s for the true file), whereas Heimdall's execution depends on \texttt{nsamps\_gulp}. Consequently, the execution time's fraction of the writer with respect to Heimdall and to the entire pipeline increases as $t_{\rm Heimdall,SIGPROC}$ decreases. 
In every considered PSRDADA case, the writer is faster than Heimdall, which means that the writer produces data faster than Heimdall GPU processing and, therefore, the pipeline is compute-bound. This means that, considering Eq.~\eqref{eq:time_real-time_pipeline}, $t_{\rm real-time} = t_{\rm Heimdall,PSRDADA}$ in any case. 
The writer-Heimdall pipeline works in this way: after Heimdall has waited for the writer to start and the writer has completed the data reading and header generation phases (see Section~\ref{sec:Tests_Writer}), the writer writes one block and Heimdall reads the same block and this procedure is completed when an entire time gulp is read in Heimdall. This writing-reading phase is very fast, involving RAM-RAM copies. This operation can require multiple iterations or a single iteration if \texttt{<b>}, \texttt{CHUNK\_SIZE}, and \texttt{nsamps\_gulp} are aligned. When a time gulp is ready, Heimdall starts processing this gulp on the GPU while the writer begins to write the data for the next gulp. Anyway, next gulp GPU processing does not start until the previous is completed, implying a waiting time in the writer between two consecutive gulps.
In the synthetic file, after a header of 327 bytes, we have a stream of $\sim$$3.88$~GiB. We now consider combinations (1)-(4), where only \texttt{nsamps\_gulp} varies. With \texttt{nsamps\_gulp} $= 2^{18} = 512$~MiB, Heimdall reads and processes about 8 gulps and $t_{\rm writer,PSRDADA} \sim 67-68\%$ of $t_{\rm Heimdall,PSRDADA}$. When \texttt{nsamps\_gulp} rises to $2^{19} = 1$~GiB and $2^{20} = 2$~GiB, Heimdall reads larger and less blocks (4 and 2, respectively), reducing Heimdall's overheads and also the waiting time of the writer, whose execution time is now $\sim$$55\%$ of $t_{\rm Heimdall,PSRDADA}$, for \texttt{nsamps\_gulp} of both $2^{19}$ and $2^{20}$. The situation drastically changes when \texttt{nsamps\_gulp} rises to $2^{21} = 4$~GiB, which is larger than the number of bytes occupied by the data in the input SIGPROC file: Heimdall processes only one gulp and the entire acquisition fits in GPU memory. Therefore, the writer only buffers data for a single gulp and then terminates, leaving Heimdall to process this single gulp on the GPU. The total execution time of the pipeline thus consists of startup operations, read/write operations for the single gulp, and a single GPU processing step. Consequently, the writer-to-Heimdall time ratio drops drastically from $\sim$55\% to $1.65-2.19$\% for all parameter combinations where \texttt{nsamps\_gulp} $= 2^{21}$ (combinations (4), (5), (6), and (8)), as the writer’s execution time remains nearly constant ($[3.271, 3.527]$ s). Slightly higher ratios occur when Heimdall runs faster (combinations (6) and (8)), while the minimum ratio ($1.65$\%) is observed in combination (4), where Heimdall is slower.

A different situation occurs when considering the true file. The data after SIGPROC header now occupy $74.69$~GiB and, therefore, this size is always larger than $Mem_{\rm nsamps\_gulp}$ (Eq.~\eqref{eq:Mem_nsamps_gulp}), leading to a number of gulps always larger than one. 
Writer-over-Heimdall time fraction maintains nearly constant with the parameter combination, passing from $86.78\%$, for (8), to $98.40\%$, for (1). Except for combination (8), all the time fractions remain between $95\%$ and $98.40\%$. Smaller time fractions are again obtained with larger \texttt{nsamps\_gulp}. For very large files, the writer tends to have a total time very similar to Heimdall's, because it spends much of its execution stuck in ring buffer synchronization operations, waiting for the consumer to consume the data. This is the behavior expected from a well-balanced producer-consumer architecture (Section~\ref{sec:Pipeline_Real-Time}), where asynchronism between the two processes is well exploited, meeting real-time constraints and offering advantages over the offline pipeline (coherently with Figures~\ref{fig:Default_Acq_Times_vs_Exec_Times_True} and~\ref{fig:Default_SIGPROC_Times_vs_PSRDADA_Times_True}). These benefits are expected to become even more pronounced once the beamformer replaces the writer in the production pipeline.

\subsubsection{Profiling}
\label{sec:Results_Performance_Profiling}

\begin{figure*}[ht]
\centering
\includegraphics[width=0.98\textwidth]{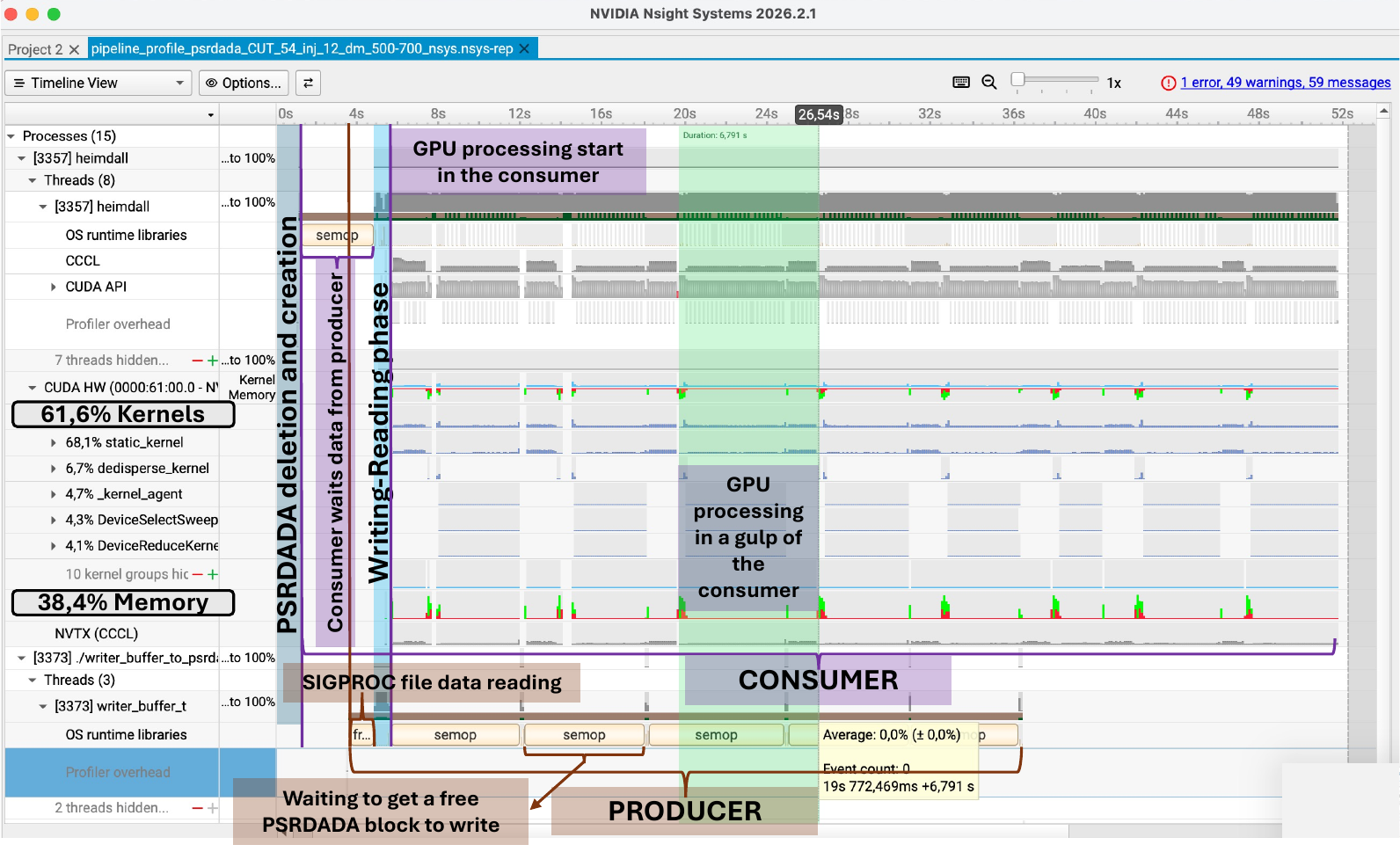}
\caption{Output of NVIDIA Nsight Systems profiler for a PSRDADA run of synthetic file n. 1 with the default combination of parameters and the restricted DM range $[500,700]$~${\rm pc}\text{ }{\rm cm}^{-3}$.}
\label{fig:Profiler}
\end{figure*}

Waiting phases of the writer and processing time of Heimdall are also visible from NVIDIA Nsight Systems profiler\footnote{\url{https://developer.nvidia.com/nsight-systems}}. Figure~\ref{fig:Profiler} illustrates one entire execution of the real-time pipeline for synthetic file n. 1 with the default parameters combination (1), considering a narrower range for DM, to avoid excessive profiling overhead. Nevertheless, the proportions among the involved processes remain the same, making the following considerations generally applicable. 
Figure~\ref{fig:Profiler} highlights in different colors some specific regions of the pipeline's execution. The dark blue blank space before the first \texttt{semop} region represents PSRDADA ring buffer deletion and creation (Section~\ref{sec:Tests_Launching_Commands}). 
The beginning of the first \texttt{semop} region (top-left part of Figure~\ref{fig:Profiler}) identifies the consumer's (Heimdall's) execution start. 
The first \texttt{semop} region indicates when Heimdall waits for the data from the producer (the writer). At a certain point of this waiting time, the writer starts its execution, 
where its first operation is the \texttt{fread}, reading the data from the SIGPROC file in the buffer (Section~\ref{sec:Tests_Writer}). After \texttt{fread}, PSRDADA header is generated but the execution time of this region is of $2\times 10^{-4}$~s, that is, negligible. The end of ``\texttt{fread} + PSRDADA header generation'' phase and the end of the first \texttt{semop} region basically coincide, since at this point data start to be available for Heimdall. The following light blue blank space identifies the ``writing PSRDADA block by the producer-reading PSRDADA block by the consumer'' phase until an entire time gulp 
is read by the consumer, as described in Section~\ref{sec:Results_Performance_Varying_Parameters}. When the first gulp is ready, core processing of Heimdall, that is, GPU processing in \texttt{hd\_execute} function, starts. 
The iterative, gulp-by-gulp execution of Heimdall is clearly visible, with a single gulp highlighted in light green. Running asynchronously, the writer executes a series of \texttt{semop} operations. These operations correspond to the time the writer spends waiting for Heimdall to consume and free a block in the PSRDADA ring buffer, which is necessary before new data can be written and the pipeline can advance.
As observed, the writer terminates its execution at $t \sim 36$~s, whereas Heimdall continues processing data until $t \sim 52$~s. The writer’s execution time amounts to $\sim$$68.5\%$ of Heimdall's, a finding consistent with the results presented in Section~\ref{sec:Results_Performance_Varying_Parameters}. Furthermore, the profiling metrics highlight the compute-bound, rather than memory-bound, nature of the pipeline: on the GPU, $61.6\%$ of the time is spent on kernel execution, compared to $38.4\%$ on memory accesses.

\section{Conclusions and Future Works}
\label{sec:Concl_Future}

The advent of next-generation radio telescopes, such as SKA, poses unprecedented computational challenges, requiring efficient solutions at the firmware, software, and hardware levels. A real-time implementation of current pipelines is an increasing impelling necessity to tackle steeply growing data-rates. The Northern Cross radio telescope is undergoing a significant upgrade thanks to the ``Next Generation – Croce del Nord'' (NG-Croce) project financed by European Union (EU) through the National Recovery and Resilience Plan (PNRR) funding scheme. The facility upgrade effort includes the refurbishment of the mechanical structure, the replacement of electrical and electronic parts with state-of-the-art technology, the installation of a new digital backend, and a HPC cluster that will allow to perform FRB searches utilizing the full instrument~\cite{Naldi_Pipeline_2025}. 
Consequently, the development of a custom, highly-optimized real-time pipeline for FRB detection is a fundamental requirement. Ongoing collaborative work with the University of Malta (as preliminarily discussed in~\cite{De_Barro_2025}) involves replacing the PSRDADA writer with the beamformer. To distribute the load, the beamformer and Heimdall will be deployed on separate, dedicated GPU nodes (Node A and Node B). PSRDADA ring buffer creation will be no more managed via command line but by a process manager, running on node B as well and acting as an orchestrator of the entire pipeline, entering in execution prior to Heimdall. Communication between the two GPU-nodes will be managed via GPU-Direct RDMA to reduce data-transfer bottlenecks.

The test pipeline presented in this work has the same logical structure as the future pipeline. Both pipelines are based on two processes, one that writes data and one that reads and analyze data. In both pipelines, the reading process is Heimdall; in the future pipeline, the writing process will be the beamformer and, in the test pipeline, is the writer. In the offline mode, the two processes are sequentially executed (Section~\ref{sec:Pipeline_Offline}); in the real-time mode, both pipelines are based on a producer-consumer approach, where the producer and the consumer are asynchronous (Section~\ref{sec:Pipeline_Real-Time}). This makes the tests presented in this paper representative for the future pipeline. Retrieving results in agreement within $\sim$3\% (TOA and DM) and $\sim$10\% (SNR) between offline and real-time modes of the test pipeline suggests the same results for the future pipeline. In terms of performance, the fact that the execution time of the offline writer is much smaller than Heimdall's execution time reduces the potential advantages of an asynchronous execution in real-time mode over a sequential execution in offline mode. However, when true files are considered and involved execution times and data sizes are larger, as expected in a realistic future scenario, the pipeline overhead is negligible compared to GPU processing, resulting in a well-balanced producer-consumer architecture where the advantages of real-time over offline processing become apparent even in this test configuration (see Figures~\ref{fig:Default_Acq_Times_vs_Exec_Times_True} and~\ref{fig:Default_SIGPROC_Times_vs_PSRDADA_Times_True}). Validation and performance tests have certainly to be repeated with the complete beamformer-Heimdall pipeline to corroborate the obtained accuracy results and to quantify the performance boost in a realistic scenario.

Furthermore, parameter space has to be further explored, 
making an analysis complementary to 
Camilleri, et al.~\cite{Camilleri_2026}, who measured the performance of the offline pipeline alone, varying a certain number of involved problem's parameters. As seen in Section~\ref{sec:Results_Performance_Varying_Parameters}, adopting certain parameter combinations allows the execution to meet real-time constraints. A proper exploration of the parameter space might lead to even more efficient combinations. Finally, the performance analysis could be explored further using tools such as the NVIDIA Nsight Compute profiler.

Future work also includes transitioning Heimdall from a single-beam to a multi-beam architecture. This step is crucial for radio transient detection, given that FRBs originate from random directions and present an escalating computational challenge as the number of antennas grows.
An open-source native multi-beam and multi-GPU version of Heimdall, with shared resources in input, already exists\footnote{\url{https://github.com/dsa110/dsa110-mbheimdall}} but has never been tested. This multi-beam version might be compared, in terms of accuracy and performance, with an embarrassingly parallel approach, running several instances of Heimdall pipeline, each processing a different beam, on different GPUs or nodes. The greatest challenge would be integrating real-time and multi-beam modes. Two further outlooks 
are the upgrade of our 
classifier FETCH with a more automated ML classification, and a compact and efficient containerization of the beamformer-Heimdall-classifier pipeline based on Apptainer\footnote{\url{https://apptainer.org/}}.

Modern HPC infrastructures\footnote{\url{https://top500.org/}}, highly-optimized codes ported to accelerators, and a real-time communication paradigm, do not only represent an increasing need in radio astronomy but also in different scientific contexts facing a steep rise of data production: ground-based experiments (e.g., CTA~\cite{CTA_Consortium_2011}); space-based experiments (e.g., Gaia~\cite{Vallenari_Gaia_Collaboration_2023} and Euclid~\cite{Laureijs_Euclid_2012}); nuclear physics experiments (e.g., LHC~\cite{Shiers_LHC_2007}; cosmological simulation codes (e.g., PLUTO~\cite{Mignone_PLUTO_2007}, RAMSES~\cite{Teyssier_RAMSES_2002}, and GADGET-4~\cite{Springel_GADGET-4_2021}). A potential use-case for the real-time paradigm might be the pipeline for covariances computation in the AVU-GSR parallel solver of the ESA Gaia mission~\cite{Becciani_Gaia_AVU-GSR_2014,Cesare_Gaia_AVU-GSR_ADASS_XXXI_2024,Cesare_INAF_Technical_Report_OpenACC_163_2022,Cesare_INAF_Technical_Report_CUDA_164_2022,Cesare_Gaia_AVU-GSR_2022_c,Cesare_Gaia_AVU-GSR_2023,Vecchiato_Gaia_AVU-GSR_ITADATA_2024_2025}. This calculation has a quadratic complexity with respect to the number of problem unknowns that will increase with the next Gaia Data Releases (DRs)~\cite{Cesare_Covariances_ADASS_XIII_2023-2025,Cesare_Covariances_SPIE_2024,Cesare_Covariances_ITADATA_2024_2025,Cesare_Covariances_PDP_2025}. Because this pipeline involves two asynchronously executed processes communicating via I/O, a real-time implementation represents a promising strategy to explore for performance improvement and storage minimization. 

\newpage

\section*{Acknowledgment}

The research activities described in this paper are carried out with the contribution of the NextGenerationEU funds within the National Recovery and Resilience Plan (PNRR), Mission 4 - Education and Research, Component 2 - From Research to Business (M4C2), Investment Line 3.1 - Strengthening and creation of Research Infrastructures, Project IR0000026 – Next Generation Croce del
Nord. 

No GenAI tool was employed in the preparation of this paper.

\bibliographystyle{IEEEtran}
\bibliography{bib}  

\end{document}